# Note on RIP-based Co-sparse Analysis

Lianlin Li

Department of Petroleum Engineering, Texas A&M University
College Station, TX, 77843

Abstract:
Over the past years, there are increasing interests in recovering the signals from undersampling data where such signals are sparse under some orthogonal dictionary or tight framework, which is referred to be *sparse synthetic model*. More recently, its counterpart, i.e., *the sparse analysis model,* has also attracted researcher's attentions where many practical signals which are sparse in the truly redundant dictionary are concerned. This short paper presents important complement to the results in existing literatures for treating sparse analysis model. Firstly, we give the natural generalization of well-known restricted isometry property (RIP) to deal with sparse analysis model, where the truly *arbitrary* incoherent dictionary is considered. Secondly, we studied the theoretical guarantee for the accurate recovery of signal which is sparse in general redundant dictionaries through solving $l_1$-norm sparsity-promoted optimization problem. This work shows not only that compressed sensing is viable in the context of sparse analysis, but also that accurate recovery is possible via solving l1-minimization problem.

## I. Introduction

Compressive sensing (CS) has become a new data acquisition theory, in which the key ingredient is the sparsity or compressibility when acquiring signals of general interest. In the nutshell of CS, the nonadaptive sampling techniques are desirable, which condenses the redudant information of compressible signal into relatively low dimensional space. In other words, far fewer measurements than unknowns are collected. By now, applications of compressed sensing are abundant and range from imaging and error correction to radar and remote sensing, see [3] and references therein.

Majority of efforts has focused on the sparse synthesis model, which has become a mature and stable field with solid theoretical foundations over long extensive study. In the context of sparse synthesis, the signal of interest $\mathbf{x}$ is *synthesized* through $\mathbf{x} = \mathbf{D}_s\boldsymbol{\alpha}$, where the vector $\boldsymbol{\alpha}$ is sparse vector, $\mathbf{D}_s \in \mathbb{R}^{n\times p}$ ( $p \geq n$ ) is some orthogonal or overcomplete dictionary, and named as synthetic operator. Therefore, the recovery of sparse signal in the context of sparse synthetic can be usually formulated as

$$\hat{\boldsymbol{\alpha}} = \arg\min_{\boldsymbol{\alpha}} \|\boldsymbol{\alpha}\|_1, \quad \text{s.t.,} \quad \|\boldsymbol{\Phi}\mathbf{D}_s\boldsymbol{\alpha} - \mathbf{y}\|_2 \leq \varepsilon \tag{P1}$$

where $\boldsymbol{\Phi} \in \mathbb{R}^{m\times n}$ ( $m < n$ ) is the so-called measurement matrix.

The counterpart one of (P1) is the so-called sparse synthetic model, where the signal of interest $\mathbf{x}$ is *analyzed* via $\boldsymbol{\alpha} = \mathbf{D}_a\mathbf{x}$ ( $\mathbf{D} \in \mathbb{R}^{p\times n}$, $p \geq n$ ), and the coefficient vector $\boldsymbol{\alpha}$ is sparse or compressible. Accordingly, the resulting optimized problem is stated in the following [1,2],

$$\hat{\mathbf{x}} = \arg\min_{\mathbf{x}} \|\mathbf{D}_a\mathbf{x}\|_1, \quad \text{s.t.,} \quad \|\boldsymbol{\Phi}\mathbf{x} - \mathbf{y}\|_2 \leq \varepsilon \tag{P2}$$

It has been shown that the solutions to (P1) and (P2) are exactly equivalent if $\mathbf{D}$ is orthongal; otherwise there is markedly different between (P1) and (P2) despite their apparent similarity [2], for example truly redundant dictionary $\mathbf{D}_a$ [2,4]. Although there are a large number of applications for (P2) with truly redundant dictionary $\mathbf{D}_a$, the compressed sensing literature is lacking on this subject.

In [2], the cosparse analysis data model as an alternative to the popular sparse synthesis model is explicitly described, where the authors pointed out that both of them are distinctly different in many cases. In this work, the authors have stated conditions that guarantee the uniqueness of cosparse solutions in the context of linear inverse problems within the framework null space analysis, and presented the efficient greedy algorithm for the cosparse recovery problem. In [1], to address (P2), the authors introduced so-called D-RIP based on the assumption of tight frame $\mathbf{D}_a$, i.e., $\mathbf{D}_a^T\mathbf{D}_a = \mathbf{I}$, and derived the strict guarantee on accurate signal recovery. However, there are numerous practical examples in which the "analysis" operator is not tight frame, but some general dictionary [2,4].

Note that in many cases the recovery results are obtained when the sensing matrix $\mathbf{\Phi}$ does not obey the restricted isometry property when the general dictionary $\mathbf{D}_a$ is considered. In this paper we desire a universal result which allows theoretical guarantee taking sensing matrix and general dictionary $\mathbf{D}_a$ into account.

## II. Main results

We now turn to discuss the generalized restricted isometry property dedicated to address the analysis-based sparse signal recovery, which can render us broader conditions about the sensing matrix under which the recovery algorithm performs well. For notational convenience, we drop the subscript of analysis operator $\mathbf{D}_a$. Analogous to well-known RIP and D-RIP recently introduced by Candes et.al., our definition of generalized RIP is following.

**Definition of Generalized RIP.** Assuming the measurement matrix $\mathbf{\Phi} \in \mathbb{R}^{m\times n}$ ($m < n$) and sparse transform matrix $\mathbf{D} \in \mathbb{R}^{p\times n}$ ($p \geq n$). $\mathbf{\Phi}$ satisfies the generalized restricted isometry property (RIP) of order *k* if there exists a $\delta_k \in (0,1)$ such that

$$(1-\delta_k)\left\|\mathbf{D}\mathbf{x}_k^D\right\|_2^2 \leq \left\|\mathbf{\Phi}\mathbf{x}_k^D\right\|_2^2 \leq (1+\delta_k)\left\|\mathbf{D}\mathbf{x}_k^D\right\| \quad (1)$$

holds for all $\mathbf{x}$ which is *k*-sparse after transformation of $\mathbf{D}$, i.e., $\|\mathbf{D}\mathbf{x}\|_0 \leq k$. Here for notable convenience, by $\mathbf{x}_k^D$ we denote $\mathbf{D}\mathbf{x}$ is *k*-sparse.

Obviously, our generalized RIP becomes the classical RIP once $\mathbf{D}=\mathbf{I}$, furthermore, we can prove that if $\mathbf{D}$ is tight frame which means $\mathbf{D}^T\mathbf{D}=\mathbf{I}$, above definition can be reduced into D-RIP introduced in [1]. There are many results that show nearly all random matrix constructions which satisfy standard RIP compressed sensing requirements will also satisfy our generalized RIP. If the matrix of $\mathbf{\Phi}$ satisfies the constraint (1), then we readily get two important conclusions as following, in particular,

**Corollary 1**.

If $\mathbf{\Phi}\in\mathbb{R}^{m\times n}$ obeys the generalized RIP represented by the inequality (1), then one has

$$\left|\left\langle \mathbf{\Phi h}_{\Lambda_i}^D, \mathbf{\Phi h}_{\Lambda_j}^D \right\rangle\right| \le \left(\delta_{2k}+\rho_k\right)\left\|\mathbf{Dh}_{\Lambda_i}^D\right\|_2 \left\|\mathbf{Dh}_{\Lambda_j}^D\right\|_2 \tag{2}$$

where $\rho_k=\max\left|\left\langle \max_{i,j}\left\langle \mathbf{Dh}_{\Lambda_i}^D, \mathbf{Dh}_{\Lambda_j}^D\right\rangle\right\rangle\right|$. In eq. (2), $\mathbf{h}_\Lambda^D$ denotes $\mathbf{h}$ is sparse in the domain $\mathbf{D}$ and its support is the support of $\Lambda$, i.e., $\Lambda=\mathbf{supp}\left(\mathbf{Dh}\right)$ together with $\left|\Lambda\right|\le k$.

**Proof.**

Firstly we normalize $\mathbf{Dh}_{\Lambda_i}^D$ and $\mathbf{Dh}_{\Lambda_j}^D$ by their 2-norm, i.e.,

$$\mathbf{Dh}_{\Lambda_i}^D \Leftarrow \frac{\mathbf{Dh}_{\Lambda_i}^D}{\left\|\mathbf{Dh}_{\Lambda_i}^D\right\|_2}, \text{ and } \mathbf{Dh}_{\Lambda_j}^D \Leftarrow \frac{\mathbf{Dh}_{\Lambda_j}^D}{\left\|\mathbf{Dh}_{\Lambda_j}^D\right\|_2}$$

Recall the identical equation

$$4\left\langle \mathbf{\Phi h}_{\Lambda_i}^D, \mathbf{\Phi h}_{\Lambda_j}^D \right\rangle = \left\|\mathbf{\Phi}\left(\mathbf{h}_{\Lambda_i}^D+\mathbf{h}_{\Lambda_j}^D\right)\right\|_2^2 - \left\|\mathbf{\Phi}\left(\mathbf{h}_{\Lambda_i}^D-\mathbf{h}_{\Lambda_j}^D\right)\right\|_2^2 \tag{3}$$

Substituting eq. (3) into eq. (1) yields to

$$4\left|\left\langle \mathbf{\Phi h}_{\Lambda_i}^D, \mathbf{\Phi h}_{\Lambda_j}^D \right\rangle\right| \le \left(1+\delta_{2k}\right)\left\|\mathbf{Dh}_{\Lambda_i}^D+\mathbf{Dh}_{\Lambda_j}^D\right\|_2^2 - \left(1-\delta_{2k}\right)\left\|\mathbf{Dh}_{\Lambda_i}^D-\mathbf{Dh}_{\Lambda_j}^D\right\|_2^2 \tag{4}$$

After using the eq. (5)

$$\left\|\mathbf{Dh}_{\Lambda_i}^D \pm \mathbf{Dh}_{\Lambda_j}^D\right\|_2^2 = \left\|\mathbf{Dh}_{\Lambda_i}^D\right\|_2^2 + \left\|\mathbf{Dh}_{\Lambda_j}^D\right\|_2^2 \pm 2\left\langle \mathbf{Dh}_{\Lambda_i}^D, \mathbf{Dh}_{\Lambda_j}^D\right\rangle = 2 \pm 2\left\langle \mathbf{Dh}_{\Lambda_i}^D, \mathbf{Dh}_{\Lambda_j}^D\right\rangle \tag{5}$$

and carrying out the anti-normalization we can arrive at inequality (2).

□

**Corollary 2.**

Suppose that $\mathbf{\Phi}$ satisfies the generalized RIP of order $2k$, and let non-zero vector $\mathbf{h}\in\mathbb{R}^n$ be arbitrary which can be represented by $\mathbf{h}=\sum_{j=0,1,...}\mathbf{h}_{\Lambda_j}^D$ together with $\Lambda_j\subset\{1,2,...,N\}$ and $|\Lambda_j|\le k$ ( $j=0,1,2,...$). Let $\Lambda_0$ be any subet of $\{1,2,...,N\}$, and $\Lambda_1$ as the index set corresponding to the $k$ entries of $\mathbf{Dh}_{\Lambda_0^c}^D$ with largest magnitude, and set $\Lambda=\Lambda_1\bigcup\Lambda_0$. Then

$$\left\|\mathbf{Dh}_{\Lambda}^D\right\|_2\le\alpha\frac{\left\|\mathbf{Dh}_{\Lambda_0^c}^D\right\|_1}{\sqrt{k}}+\beta\frac{\left|\left\langle\mathbf{\Phi h}_{\Lambda}^D,\mathbf{\Phi h}\right\rangle\right|}{\left\|\mathbf{Dh}_{\Lambda}^D\right\|_2} \tag{6}$$

with $\alpha=\dfrac{\sqrt{2}\left(\delta_{2k}+\rho\right)}{1-\delta_{2k}}$ 和 $\beta=\dfrac{1}{1-\delta_{2k}}$ 。

**Proof.**

Firstly consider the expansion

$$\begin{aligned}\left\|\mathbf{\Phi h}_{\Lambda}^D\right\|_2^2&=\left\langle\mathbf{\Phi h}_{\Lambda}^D,\mathbf{\Phi h}\right\rangle-\left\langle\mathbf{\Phi h}_{\Lambda}^D,\mathbf{\Phi h}_{\Lambda^c}^D\right\rangle\\&=\left\langle\mathbf{\Phi h}_{\Lambda}^D,\mathbf{\Phi h}\right\rangle-\sum_{j\ge2}\left\langle\mathbf{\Phi h}_{\Lambda}^D,\mathbf{\Phi h}_{\Lambda_j}^D\right\rangle\end{aligned} \tag{7}$$

Using Eq. (2) we can get the bound estimation as following

$$\begin{aligned}\left|\sum_{j\ge2}\left\langle\mathbf{\Phi h}_{\Lambda}^D,\mathbf{\Phi h}_{\Lambda_j}^D\right\rangle\right|&\le\sum_{j\ge2}\left|\left\langle\mathbf{\Phi h}_{\Lambda}^D,\mathbf{\Phi h}_{\Lambda_j}^D\right\rangle\right|\\&\le\sum_{j\ge2}\left|\left\langle\mathbf{\Phi h}_{\Lambda_0}^D,\mathbf{\Phi h}_{\Lambda_j}^D\right\rangle\right|+\sum_{j\ge2}\left|\left\langle\mathbf{\Phi h}_{D\Lambda_1},\mathbf{\Phi h}_{\Lambda_j}^D\right\rangle\right|\\&\le\left(\delta_{2k}+\rho\right)\left(\left\|\mathbf{Dh}_{\Lambda_0}^D\right\|_2+\left\|\mathbf{Dh}_{\Lambda_1}^D\right\|_2\right)\sum_{j\ge2}\left\|\mathbf{Dh}_{\Lambda_j}^D\right\|_2\\&\le\frac{\sqrt{2}\left(\delta_{2k}+\rho\right)\left\|\mathbf{Dh}_{\Lambda}^D\right\|_2\left\|\mathbf{Dh}_{\Lambda_0^c}^D\right\|_1}{\sqrt{k}}\end{aligned} \tag{8}$$

where $\left\|\mathbf{Dh}_{\Lambda_0}^D\right\|_2+\left\|\mathbf{Dh}_{\Lambda_1}^D\right\|_2\le\sqrt{2}\left\|\mathbf{Dh}_{\Lambda}^D\right\|_2$ and $\sum_{j\ge2}\left\|\mathbf{Dh}_{\Lambda_j}^D\right\|_2\le\frac{1}{\sqrt{k}}\left\|\mathbf{Dh}_{\Lambda_0^c}^D\right\|_1$ have been used.

Combing eq. (7) and (8) we can obtain the following conclusion, i.e.,

$$
\begin{aligned}
\left\|\mathbf{\Phi h}_{\Lambda}^{D}\right\|_2^2 &= \left\langle \mathbf{\Phi h}_{\Lambda}^{D}, \mathbf{\Phi h}\right\rangle - \sum_{j\geq 2}\left\langle \mathbf{\Phi h}_{\Lambda}^{D}, \mathbf{\Phi h}_{\Lambda_j}^{D}\right\rangle \\
&\leq \left|\left\langle \mathbf{\Phi h}_{\Lambda}^{D}, \mathbf{\Phi h}\right\rangle\right| + \left|\sum_{j\geq 2}\left\langle \mathbf{\Phi h}_{\Lambda}^{D}, \mathbf{\Phi h}_{\Lambda_j}^{D}\right\rangle\right| \\
&\leq \left|\left\langle \mathbf{\Phi h}_{\Lambda}^{D}, \mathbf{\Phi h}\right\rangle\right| + \frac{\sqrt{2}\left(\delta_{2k}+\rho\right)\left\|\mathbf{Dh}_{\Lambda}^{D}\right\|_2\left\|\mathbf{Dh}_{\Lambda_0^c}^{D}\right\|_1}{\sqrt{k}}
\end{aligned}
\tag{9}
$$

Using eq. (9) in eq. (1) we can get the upper bound of $\left\|\mathbf{Dh}_{\Lambda}^{D}\right\|_2$, i.e.,

$$
\left\|\mathbf{Dh}_{\Lambda}^{D}\right\|_2 \leq \frac{\dfrac{\left|\left\langle \mathbf{\Phi h}_{\Lambda}^{D}, \mathbf{\Phi h}\right\rangle\right|}{\left\|\mathbf{Dh}_{\Lambda}^{D}\right\|_2} + \dfrac{\sqrt{2}\left(\delta_{2k}+\rho\right)\left\|\mathbf{Dh}_{\Lambda_0^c}^{D}\right\|_1}{\sqrt{k}}}{1-\delta_{2k}}
\tag{10}
$$

which close the proof of corollary 2.

□

We now turn to the discussion of the reconstruction of co-sparsity promoted by l1-norm. Similar as done in [1,2,4] and others, this problem can be formulated into

$$
\hat{\mathbf{x}} = \arg\min_{\mathbf{z}} \left\|\mathbf{Dz}\right\|_1, \quad \text{s.t.,} \quad \mathbf{z}\in\mathcal{B}\left(\mathbf{y}\right)
$$

With above armed, we can state our main result summarized in theorem 1, i.e.,

**Theorem 1.**

Suppose that $\mathbf{\Phi}$ satisfies the generalized RIP of order $2k$ where $\delta_{2K} < \sqrt{2}-1$, and let non-zero vector $\mathbf{h}=\hat{\mathbf{x}}-\mathbf{x}$ ( $\hat{\mathbf{x}},\mathbf{x}\in\mathbb{R}^N$ ) which can be represented by $\mathbf{h}=\sum_{j=0,1,\ldots}\mathbf{h}_{\Lambda_j}^{D}$ together with $\Lambda_j\subset\{1,2,\ldots,N\}$ and $\left|\Lambda_j\right|\leq k$ ( $j=0,1,2,\ldots$). Let $\Lambda_0$ the index set corresponding to the $k$ entries of $\mathbf{Dx}$ with largest magnitude, and $\Lambda_1$ as the index set corresponding to the $k$ entries of $\mathbf{Dh}_{\Lambda_0^c}^{D}$ with largest magnitude, and set $\Lambda=\Lambda_1\cup\Lambda_0$. If $\left\|\mathbf{D}\hat{\mathbf{x}}\right\|_1\leq\left\|\mathbf{Dx}\right\|_1$, then

$$
\left\|\mathbf{Dh}\right\|_2 \leq C_0\frac{\sigma_k\left(\mathbf{x}\right)}{\sqrt{k}} + C_1\frac{\left|\left\langle \mathbf{\Phi h}_{\Lambda}^{D}, \mathbf{\Phi h}\right\rangle\right|}{\left\|\mathbf{Dh}_{\Lambda}^{D}\right\|_2}
\tag{11}
$$

其中 $C_0 = 2\dfrac{1-\left(1-\sqrt{2}\right)\delta_{2k}}{1-\left(1+\sqrt{2}\right)\delta_{2k}}$， $C_0 = \dfrac{2}{1-\left(1+\sqrt{2}\right)\delta_{2k}}$ 。

**Proof.**

Strictly along the line used in [3], we can readily finish the proof of theorem 1. Starting from and fact of $\mathbf{h} = \mathbf{h}_{D\Lambda} + \mathbf{h}_{D\Lambda^c}$ and the triangle inequality, we have

$$\|\mathbf{Dh}\|_2 \le \left\|\mathbf{Dh}_{\Lambda}^{D}\right\|_2 + \left\|\mathbf{Dh}_{\Lambda^c}^{D}\right\|_2 \le \left\|\mathbf{Dh}_{\Lambda}^{D}\right\|_2 + \frac{\left\|\mathbf{Dh}_{\Lambda_0^c}^{D}\right\|_1}{\sqrt{k}} \tag{12}$$

where $\left\|\mathbf{Dh}_{\Lambda^c}^{D}\right\|_2 \le \frac{\left\|\mathbf{Dh}_{\Lambda_0^c}^{D}\right\|_1}{\sqrt{k}}$ have been used.

Since $\|\mathbf{D\hat{x}}\|_1 \le \|\mathbf{Dx}\|_1$ by applying the triangle inequality we obtain

$$\begin{aligned} \|\mathbf{Dx}\|_1 &\ge \|\mathbf{Dx} + \mathbf{Dh}\|_1 \\ &= \left\|\mathbf{Dx}_{\Lambda_0}^{D} + \mathbf{Dh}_{\Lambda_0}^{D}\right\|_1 + \left\|\mathbf{Dx}_{\Lambda_0^c}^{D} + \mathbf{Dh}_{\Lambda_0^c}^{D}\right\|_1 \\ &\ge \left\|\mathbf{Dx}_{\Lambda_0}^{D}\right\|_1 - \left\|\mathbf{Dh}_{\Lambda_0}^{D}\right\|_1 + \left\|\mathbf{Dh}_{\Lambda_0^c}^{D}\right\|_1 - \left\|\mathbf{Dx}_{\Lambda_0^c}^{D}\right\|_1 \end{aligned}$$

which means

$$\left\|\mathbf{Dh}_{\Lambda_0^c}^{D}\right\|_1 \le 2\left\|\mathbf{Dx}_{\Lambda_0^c}^{D}\right\|_1 + \left\|\mathbf{Dh}_{\Lambda_0}^{D}\right\|_1 = \left\|\mathbf{Dh}_{\Lambda_0}^{D}\right\|_1 + 2\sigma_k(\mathbf{x}) \tag{13}$$

where $\sigma_K(\mathbf{x}) = \left\|\mathbf{Dx}_{\Lambda_0^c}^{D}\right\|_1$ 。

Using above inequality in $\left\|\mathbf{Dh}_{\Lambda^c}^{D}\right\|_2 \le \frac{\left\|\mathbf{Dh}_{\Lambda_0^c}^{D}\right\|_1}{\sqrt{k}}$ yields to

$$\left\|\mathbf{Dh}_{\Lambda^c}^{D}\right\|_2 \le \frac{\left\|\mathbf{Dh}_{\Lambda}^{D}\right\|_1 + 2\sigma_k(\mathbf{x})}{\sqrt{k}} \le \left\|\mathbf{Dh}_{\Lambda}^{D}\right\|_2 + 2\frac{\sigma_k(\mathbf{x})}{\sqrt{k}} \tag{14}$$

Combing equations (13) and (14) we have

$$\|\mathbf{Dh}\|_2 \le 2\left\|\mathbf{Dh}_{\Lambda}^{D}\right\|_2 + 2\frac{\sigma_k(\mathbf{x})}{\sqrt{k}} \tag{15}$$

On the other hand, from corollary 2 we know that

$$
\begin{aligned}
\left\|\mathbf{Dh}_{\Lambda}^{D}\right\|_2 &\le \alpha \frac{\left\|\mathbf{Dh}_{\Lambda_0^c}^{D}\right\|_1}{\sqrt{K}} + \beta \frac{\left|\left\langle \mathbf{\Phi h}_{\Lambda}^{D}, \mathbf{\Phi h}\right\rangle\right|}{\left\|\mathbf{Dh}_{\Lambda}^{D}\right\|_2} \\
&\le \alpha \frac{\left\|\mathbf{Dh}_{\Lambda_0}^{D}\right\|_1 + 2\sigma_k(\mathbf{x})}{\sqrt{k}} + \beta \frac{\left|\left\langle \mathbf{\Phi h}_{\Lambda}^{D}, \mathbf{\Phi h}\right\rangle\right|}{\left\|\mathbf{Dh}_{\Lambda}^{D}\right\|_2} \\
&\le \alpha \left\|\mathbf{Dh}_{\Lambda_0}^{D}\right\|_2 + 2\alpha \frac{\sigma_k(\mathbf{x})}{\sqrt{k}} + \beta \frac{\left|\left\langle \mathbf{\Phi h}_{\Lambda}^{D}, \mathbf{\Phi h}\right\rangle\right|}{\left\|\mathbf{Dh}_{\Lambda}^{D}\right\|_2} \\
&\le \alpha \left\|\mathbf{Dh}_{\Lambda}^{D}\right\|_2 + 2\alpha \frac{\sigma_k(\mathbf{x})}{\sqrt{k}} + \beta \frac{\left|\left\langle \mathbf{\Phi h}_{\Lambda}^{D}, \mathbf{\Phi h}\right\rangle\right|}{\left\|\mathbf{Dh}_{\Lambda}^{D}\right\|_2}
\end{aligned} \tag{16}
$$

which means

$$
\left\|\mathbf{Dh}_{\Lambda}^{D}\right\|_2 \le \frac{2\alpha}{1-\alpha}\frac{\sigma_k(\mathbf{x})}{\sqrt{k}} + \frac{\beta}{1-\alpha}\frac{\left|\left\langle \mathbf{\Phi h}_{\Lambda}^{D}, \mathbf{\Phi h}\right\rangle\right|}{\left\|\mathbf{Dh}_{\Lambda}^{D}\right\|_2} \tag{17}
$$

Substituting equation (17) into (12), we obtain the upper bound of $\left\|\mathbf{Dh}\right\|_2$ as

$$
\begin{aligned}
\left\|\mathbf{Dh}\right\|_2 &\le 2\left(\frac{2\alpha}{1-\alpha}\frac{\sigma_k(\mathbf{x})}{\sqrt{k}} + \frac{\beta}{1-\alpha}\frac{\left|\left\langle \mathbf{\Phi h}_{\Lambda}^{D}, \mathbf{\Phi h}\right\rangle\right|}{\left\|\mathbf{Dh}_{\Lambda}^{D}\right\|_2}\right) + 2\frac{\sigma_k(\mathbf{x})}{\sqrt{k}} \\
&\le \left(\frac{4\alpha}{1-\alpha} + 2\right)\frac{\sigma_k(\mathbf{x})}{\sqrt{k}} + \frac{2\beta}{1-\alpha}\frac{\left|\left\langle \mathbf{\Phi h}_{\Lambda}^{D}, \mathbf{\Phi h}\right\rangle\right|}{\left\|\mathbf{Dh}_{\Lambda}^{D}\right\|_2}
\end{aligned}
$$

which completes the proof of theorem 1.

Remark:

For different setup of $\mathcal{B}(\mathbf{y})$ for example,

noise-free case $\mathcal{B}(\mathbf{y}) = \{\mathbf{z}, \mathbf{\Phi z} = \mathbf{y}\}$,

noisy observation $\mathcal{B}(\mathbf{y}) = \left\{\mathbf{z}, \left\|\mathbf{\Phi z} - \mathbf{y}\right\|_2 \le \varepsilon\right\}$

and

*Dantzig* selector $\mathcal{B}(\mathbf{y}) = \left\{\mathbf{z}, \left\|\mathbf{\Phi}^T(\mathbf{\Phi z} - \mathbf{y})\right\|_\infty \le \lambda\right\}$,

from theorem 1 we can straightforward derive the corresponding theoretical guarantees for the stable recovery along the almost same line as used in [3]. Due to space limitation, we leave this part for the reader.

□

## III. Conclusion

This short paper presents general result of recovering sparse signals which are sparse in the truly redundant dictionary, which complements to the results in existing literatures for treating analysis based sparse recovery. To end this, we give the natural generalization of well-known restricted isometry property (RIP) to deal with the recovery of signal sparse in some *arbitrary* incoherent dictionary. Afterwards, we studied the theoretical guarantee for the accurate sparse recovery of signal which is sparse in highly arbitrary overcomplete and coherent dictionaries through solving $l_1$-norm sparsity-promoted optimization problem.